# UXer-AI Collaboration Process for Enhancing Trust

Author: Harin Yoon, Dongwhan Kim, Changhoon Oh, Soojin Jun

## Abstract

In recent years, discussions on integrating Artificial Intelligence (AI) into UX design have intensified. However, the practical application of AI tools in design is limited by their operation within overly simplified scenarios, inherent complexity and unpredictability, and a general lack of relevant education. This study proposes an effective UXer-AI collaboration process to address these issues and seeks to identify efficient AI collaboration strategies through a series of user studies. In a preliminary study, two participatory design workshops identified major barriers to UXer-AI collaboration, including unfamiliarity with AI, inadequate internal support, and trust issues. To address the particularly critical issue of diminished trust, this study developed a new AI prototype model, TW-AI, that incorporates verification and decision-making processes to enhance trust and operational efficiency in UX design tasks. Task performance experiments and in-depth interviews evaluated the TW-AI model, revealing significant improvements in practitioners' trust, work efficiency, understanding of usage timing, and controllability. The "Source" function, based on Retrieval-Augmented Generation (RAG) technology, notably enhanced the reliability of the AI tool. Participants noted improved communication efficiency and reduced decision-making time, attributing these outcomes to the model's comprehensive verification features and streamlined approach to complex verification tasks. This study advances UXer-AI collaboration by providing key insights, bridging research and practice with actionable strategies, and establishing guidelines for AI tool designs tailored to UX. It contributes to the HCI community by outlining a scalable UXer-AI collaboration framework that addresses immediate operational challenges and lays the foundation for future advancements in AI-driven UX methodologies.

# 1. INTRODUCTION

In recent years, there has been extensive discourse on the integration of artificial intelligence (AI) into user experience (UX) design (Abbas et al., 2022; Gmeiner et al., 2023; Li et al., 2024; Lu et al., 2022). Despite academic interest and research, AI tools supporting UX design have yet to significantly impact the industry, with minimal adoption by practitioners (Lu et al., 2022; Shi et al., 2023). This gap arises because AI tools are designed to operate only within overly simplified scenarios, and the main focus of research on AI utilization has been merely on introducing new technologies (Shi et al., 2023). Additionally, the complexity of AI technology, the unpredictability of outcomes, the lack of relevant education, and the rapid pace of technological change make it difficult for most designers to properly utilize AI (Dove et al., 2017; Gmeiner et al., 2023; Sun et al., 2020; Yang et al., 2018; Yang et al., 2020). To bridge this gap, this study aims to integrate collaborative tasks with generative AI into the UX design process. Achieving this requires a thorough understanding of how UX designers interact with AI and the challenges they face in this process. In this regard, Stige et al. (2023) highlighted the fragmented nature of discussions on integrating AI into the UX design process and emphasized the need for research on the potential negative impacts of AI in UX design work. Therefore, this research seeks to investigate the obstacles UX designers encounter when collaborating with AI. Specifically, our goal is to propose a collaborative process and AI prototype model that UX designers can apply in practice, alongside the collaboration process between UX design practitioners and AI.

The research was conducted as follows: First, we reviewed the literature on relevant processes, including the double diamond process, and then defined the UX practitioners(UXers) and AI collaboration process. Subsequently, in the preliminary study, two participatory design workshops were conducted, with participants divided based on their professional experience, to gather opinions from a total of 42 UXers. The results of the workshops were synthesized to identify obstacles to UXer-AI collaboration and to compare practitioners' perceptions based on their experience levels, which then informed the formulation of design goals. In line with the design goals, the UXer-AI collaboration process was revised, and a generative AI prototype following the new process was created. To verify the effectiveness of the revised collaboration process and prototype, task execution observation experiments and in-depth interviews were conducted with 20 practitioners with UX design-related job experience. The results included a comparison of trust scores between the existing generative AI and the prototype AI tools and an evaluation of the effectiveness of the revised process and prototype. Additionally, based on the interview content, the discussion addressed four factors influencing trust and the effectiveness of the AI prototype, as well as insights related to the competencies and roles required for UXers in the era of artificial intelligence.

This paper contributes to the human-computer interaction (HCI) field in the following ways. First, by conducting participatory design workshops with UX practitioners, it provides insights that can bridge the gap between research and practice, enhancing the understanding of practitioners' perceptions of AI in practical settings. Second, a collaboration process and an AI prototype were proposed to help UXers collaborate more effectively and efficiently with AI. This prototype is the result of implementing existing generative AI technology in a way that alleviates key barriers to UXer-AI collaboration. Third, the research process provided researchers and practitioners with the opportunity to consider how the roles and necessary competencies of UXers will change in the future. The UXer-AI collaboration process carried out throughout the study demonstrates that outcomes vary depending on how AI is utilized, highlighting the importance of UXers' capabilities and methods in leveraging AI.

## 2. RELATED WORK

### 2.1 UXer-AI Collaboration
#### 2.1.1 Collaborative Relationship Between UXers and AI

In the field of UX design, there has been discussion about the potential for AI to replace human designers (Li et al., 2023). Consequently, it has been revealed that many designers treat AI as a magic tool and tend to blindly trust it (Dove et al., 2017). However, experts predict that within the next decade, nearly all humans and AIs will work in teams (Fischer, 2023), and due to areas where AI cannot perform, the likelihood of human designers being replaced is considered low (Boni, 2021; Heer, 2019; Kaiser, 2019; Nazli, 2022; Sendhoff & Wersing, 2020). Moreover, an analysis of over 1,600 cases of AI application in more than 45 fields showed that a collaborative relationship with AI, rather than a competitive one, would hugely benefit both parties (O'Donovan et al., 2015; Wu et al., 2021). For example, designers can enhance their creativity by generating more diverse design prototypes through collaboration with AI (Dove et al., 2017). Considering this, AI should not replace humans but instead enhance human capabilities (Xu, 2019). If UXers were to adopt this human-centered approach, AI tools could be used more ethically and effectively, thus avoiding unintended harmful outcomes (Bond et al., 2019). Therefore, this study aimed to treat AI not as a replacement for human UX designers but as a collaborative partner to perform tasks more efficiently.

The term "human in the loop" (HITL), which denotes a person guiding an AI system's learning within its operational processes to ensure more stable outcomes through training, testing, or adjustments, was adapted in this study within the context of UX design to "UXer in the loop" (UITL). This concept restricts the scope to UX designers as collaborative agents, specifically referring to them as involved in demonstrating empathy and understanding within the AI collaboration process to generate better outcomes uniquely achievable by human capabilities. The UITL fundamentally trains data to enable AI to produce appropriate outcomes and continuously adjusts AI roles and operations to fit task-specific UX processes (Shi et al., 2023). Additional roles may be necessary for the UITL to collaborate effectively with AI. Therefore, this research analyzed the UXer-AI collaboration process to derive implications regarding the necessary capabilities and roles of the UITL. Subsequent sections consistently refer to UITL as the UXer, meaning a collaborative agent whose roles and definitions align with those described in this section.

#### 2.1.2 Derivation of the UXer-AI Collaboration Process

*Table 1. Related Process*

| Process | Contents | Reference |
|---|---|---|
| Double diamond process | Discover - Define - Develop - Deliver | Design Council, 2004 |
| Design thinking process | Empathize - Define - Ideate - Prototype - Test | Instiute of Design at Stanford, 2024 |
| UX design process | Analyze - Design - Implement - Evaluate | Hartson and Pyla, 2012 |
| Collaborative creativity model | Focus - Frame - Create - Complete | Aragon and Williams, 2011 |
| Collaborative designing model | Establish goals - Search problem space - Determine constraints - Construct solution | Lahti et al., 2004 |
| Human-AI co-creation model | Perceive - Think - Express – Collaborate - Build - Test | Wu et al., 2021 |

A review of six processes, including UX design processes and AI-human collaboration processes (Table 1), revealed that all share a common flow of problem identification, solution generation, and output creation processes in sequential order. Considering this, it can be anticipated that the collaboration process between UXers and AI will also follow a similar three-step sequence. One distinctive aspect of the collaboration process with AI compared to other processes is the iterative nature of prompt input and output-creation tasks at each stage. This consideration stems from the flexible nature of generative AI, defined in this paper as a collaborative agent capable of generating new outputs based on user input (Shi et al., 2023). Thus, the process is expected to have a cyclical nature where humans input, AI generates, and humans input again until final outcomes are constructed and tested.

In the preliminary study, workshops were conducted where UXers directly participated in design tasks to comprehensively identify factors hindering UXer-AI collaboration. Based on these findings, this study aimed to propose and evaluate effective collaboration strategies and AI tool prototypes to mitigate hindering factors, thereby enhancing UXer-AI collaboration experiences and improving the efficiency of UX design workflow processes.

**2.2 Challenges for UXers and AI Collaboration**
        **2.2.1 Lack of Trust in AI**

The pivotal factor affecting the integration of AI into practical use is user trust in AI technology (Glikson & Woolley, 2020). Users wish to believe that AI will function correctly, to understand its operation, and to know how to use it appropriately. Hence, the importance of trust in interactions with AI has been emphasized by much prior research (Sullivan et al., 2022). As trust is a core element addressed within AI ethics, it is essential to understand UXers' trust in AI and discuss strategies to enhance trust for effective UXer-AI collaboration.

Trust is formed based on predictability (Rempel et al., 1985) and can evolve over time, though it tends to be informed more by risks than potential benefits, emphasizing the importance of initial experiences as a trustworthy starting point (Penny Collisson, 2015, as cited in Wang & Moulden, 2021). Disappointing or unreliable experiences with AI can weaken trust, negatively impacting the continuous use of the technology (Wang & Moulden, 2021). Therefore, understanding how users come to trust AI is essential. Factors influencing AI trust are defined differently by researchers but commonly include considerations of fairness, accountability, reliability, and transparency (Chintada, 2021; Glikson & Woolley, 2020; Omrani et al., 2022; Sison et al., 2023).

At the core of trust is reducing complexity and uncertainty (Muir, 1994). In different fields where AI is used, the causes of complexity and uncertainty can vary (Lankton et al., 2015; Omrani et al., 2022), and the four factors affecting trust may also operate differently. Therefore, in the preliminary study, the comprehensive identification of factors hindering UXer-AI collaboration, including the four elements of trust, was carried out.

Specifically, the study compared UXers' perceptions of trust in AI across different career stages, proposed and validated strategies for enhancing trustworthiness, and conducted in-depth analysis of factors contributing to trust deterioration.

### 2.2.2 Absence of AI Tools for UX Design

A trustworthy human-AI collaborative relationship entails leveraging each other's strengths, mitigating weaknesses, and assuming responsibility for outcomes (Ramchurn et al., 2021), these principles that applies to UXer-AI collaboration as well. However, due to the intricate and often opaque nature of AI processes, most designers struggle to effectively control and pragmatically utilize AI (Yang et al., 2020). Moreover, AI tools supporting design lack real-time communication and cooperative capabilities (Ryu et al., 2021), creating a lack of suitable AI tools for collaboration with UXers. In reviewing 50 AI-based design tools, a study found that only 14 were relevant to UX/UI design (Ryu et al., 2021), and even those tools primarily focused on user interface (UI) design, with a special emphasis on prototyping (Pandian & Suleri, 2020). Consequently, many AI tools supporting UX design currently fail to make a substantial impact in practical settings (Lu et al., 2023; Shi et al., 2023). Therefore, developing AI tools tailored for UX design practice is crucial to foster effective UXer-AI collaboration, necessitating a review of design strategies and considerations.

The factors determining the characteristics of designer-AI collaboration, as proposed by Shi et al. (2023), include scope, access, agency, flexibility, and visibility. These attributes can aid in concretizing the design of AI tools. First, it is essential to define the range of tasks that can be conducted through UXer-AI collaboration. This study identified the stages where UXer-AI collaboration can be most effectively achieved through two workshops and proposes an AI model focusing on those stages. Furthermore, depending on the practical experience of UX practitioners, the reliance on specialized knowledge and skills may vary. As such, these differences were analyzed to derive design implications for AI models that all UX practitioners can utilize. Regarding agency, the features of UXer-AI collaboration may vary depending on who takes the lead. Predictably, the collaborative process defined in this study suggests that UXers will have a designer-led characteristic, where they make all decisions throughout the process, as they need to iteratively adjust inputs to meet given design requirements. Additionally, if UXers can review or modify outputs generated by AI in real time, flexibility in collaboration is enhanced. Furthermore, if UXers can clearly perceive that they are collaborating with AI, explicit collaboration becomes feasible. Taking into account these five factors, this study aimed to propose a design strategy for AI tools that pursues a human-centered approach to collaboration under the direction of UXers.

## 3. PRELIMINARY STUDY[1]
### 3.1 Study Procedure
To observe UX practitioners working with AI tools in executing real-world tasks and to gain in-depth insights into the challenges UX practitioners face, we conducted a participatory design workshop using generative AI. The specific research questions were as follows:

---

[1] This chapter is based on the following paper: Yoon, H., Oh, C., & Jun, S. (2024, May). How Can I Trust AI?: Extending a UXer-AI Collaboration Process in the Early Stages. In Extended Abstracts of the CHI Conference on Human Factors in Computing Systems (pp. 1-7).

1) What are the main barriers for UXers in the UXer-AI collaboration process?
2) What are the design goals for the AI tool to lower the barriers to UXer-AI collaboration?

Table 2. Workshop (Practical Exercise on UX Design Utilizing Generative AI) Procedures

| Participants | 27 UXers (6 or more years of experience) | | |
|---|---|---|---|
| Steps | Education (60 minutes) | Task Performance (180 minutes) | In-Depth Interview |
| Contents | Introducing concepts and usage methods to enhance the understanding of generative AI | 1. Recognize problem (text-to-text AI) 2. Derive solutions (text-to-text AI) 3. Generate the final outcome (text-to-image AI) | 1. Difficulties 2. Potential applications in practice 3. Concerns |

The workshop included UX design professionals with 6 to 37 years of experience and was conducted in three phases. First, a pre-training session was held to accommodate participants unfamiliar with AI. This was followed by tasks involving the use of eight different generative AIs. Finally, participants were divided into four groups of 6-7 members each for focus group interviews about their AI usage experiences. The tasks were structured to follow the three-step UXer-AI collaboration process (problem identification, solution generation, and output creation), with the goal being to redesign the Netflix mobile UI. This task objective was selected because the generative AI had already been trained on sufficient data on this service, and it was a service that participants were familiar with and could relate to.

Eight different AI tools were used during the task including text-generation AIs such as ChatGPT and Bard, image-generation AIs such as Stable Diffusion and Blue Willow, and prototype-generation AIs such as Figma AI. During the task, participants shared their opinions on AI collaboration and the outcomes of each task phase in real time using the collaborative tool Padlet.

3.2 Data Analysis

The data collected from participants' comments during the workshop, their inputs on Padlet, and the notes transcribed from the focus group interviews were analyzed using thematic analysis. We utilized Atlas.ti, a qualitative data analysis tool, and followed a process consisting of three steps: initial open coding, reviewing initial codes and categorizing them into higher-level themes, and refining subcategories. As a result, the data were organized into a total of 81 initial codes, 27 sub-themes, and finally, 10 major categories. Focusing on the research questions, we derived insights based on the eight major categories that represent the barriers to UXer-AI collaboration (Table 3).

3.3 Factors that Hinder UXer-AI Collaboration

Table 3. Eight Factors that Hinder UXer–AI Collaboration

| Category | Element | Content |
| --- | --- | --- |
| Challenges arising from unfamiliarity with AI | Prompt Input | "In the end, how you use the prompt is the most crucial. Using the prompt is challenging." |
| | Generate | "Even when using GPT-4, the specificity remains limited." "I'm unsure how my requirements have been reflected on the screen." |
| | Tool selection | The participants, before undertaking the task, asked, "Which tools are best to use?" After completion, they discussed the optimal combination of programs. |
| Issues requiring internal support within the company | Financial | "Since all the truly usable tools are behind a paywall, utilizing a paid AI for the task would have been more beneficial." |
| | Security | When using the internal corporate network, there is an issue with access to a specific AI service being blocked. |
| Critical barriers in UX design tasks relying on research data | Hallucination | "Since there is no back data provided for the answers, a verification process is needed to ensure they are not hallucinations." |
| | Reliability | "GPT doesn't provide sources, so how can we verify this?" |
| | Decision-making | "In the absence of source data, it becomes necessary to perform tasks like A/B testing to filter out trustworthy responses." |

The primary barriers to UXer-AI collaboration identified in the study included challenges due to unfamiliarity with AI, issues requiring internal company support, and barriers critical to UX design work.

First, challenges due to unfamiliarity with AI encompass difficulties in knowing how to input prompts correctly and how to generate desired outcomes, as well as uncertainties about which AI tools to use among various options. These difficulties were observed to gradually diminish as participants became more accustomed with AI through repeated task performance.

Next, concerns about AI usage fees and security issues were categorized as problems requiring internal company support. To effectively use AI in a practical environment, support from a company is essential.

Lastly, the most significant factors hindering UXer-AI collaboration were the AI hallucination problem and the resulting decrease in trustworthiness due to the lack of supporting evidence. The inability to trust AI-generated outputs led to difficulties in making decisions. Since UX design relies on generating outcomes based on researched data, UXers' inability to trust AI is one of the most significant challenges to effective collaboration with AI.

### 3.4 Additional Workshop: Comparing Trust Levels in AI Based on Practical Experience

The results of the first workshop confirmed that the biggest barrier to UXer-AI collaboration is the lack of trust in AI-generated outcomes. Therefore, the second workshop centered on the level of trust in AI. Since perceptions of and trust in technology can vary depending on experience, the objective was to compare these differences.

Participants with less than 3 years of experience were selected to elucidate the differences with those from the first workshop. The workshop procedures were identical to those of the first workshop. However, it was anticipated that junior practitioners would require more detailed explanations and guidance during the practical exercises using AI for UX design tasks, so participation was limited to 15 individuals. As a result of a smaller number of participants, the total workshop duration was reduced by 90 minutes compared to the workshop for experienced practitioners. The workshop allocated 60 minutes to education to enhance the understanding of generative AI, 120 minutes to performing UX design tasks using various generative AI tools, and 30 minutes to a focus group interview.

### 3.5 Design Goals Considering Differences in AI Trust Levels

Contrary to the senior practitioners who commonly pointed out hallucination and reliability issues, junior practitioners' responses were divided into two categories. The difference in trust levels in AI between these individuals with different experience levels is summarized in Table 4.

*Table 4. Comparison of Trust Levels Based on UX Design Work Experience*

| Group | Trust Level | Description |
| --- | --- | --- |
| Junior Group | Relatively Low Trust | Concerned about hallucinations but found separate verification tasks cumbersome |

|  | High Trust | Judged the generated responses as persuasive and rated their trustworthiness highly |
|---|---|---|
| Senior Group | Low Trust | Questioned the reliability of the data and raised the need for verification tasks |

Given the differences in the perceptions of and trust in AI of practitioners with different experience levels identified through the two workshops, the following three design goals were derived.

**Design Goal 1: A system where all users can perform verification tasks regardless of their awareness of hallucination.**

Senior practitioners, while utilizing AI for UX research, pointed out the absence of evidence and emphasized the need for verification. They demonstrated awareness of hallucination issues and consciously sought to verify responses. In contrast, junior practitioners often lacked familiarity with the concept of hallucination or, even if aware, did not recognize the need for additional verification. Considering the research findings that suggest increased cognitive trust when AI is perceived as human-like (Omrani et al., 2022), there is a risk that UXers may uncritically trust and accept responses generated by conversational AI that mimics human-like interactions. Some junior practitioners even evaluated AI's reliability highly without recognizing hallucination issues. Furthermore, considering that junior practitioners mentioned the cumbersome nature of verification tasks despite their awareness of hallucination issues, they are likely to overlook the need to validate responses. Since UX design involves deriving final outcomes based on research findings, failure to verify AI responses could lead to significant issues. Therefore, it is essential in practice to review the factual accuracy of research data and related evidence. Consequently, there is a need for designs that facilitate verification tasks for all users, regardless of their awareness of hallucination issues.

**Design Goal 2: A system that integrates multiple verification methods based on current available technologies, including source citation and web search for answers.**

Despite recognizing the need for verification, senior practitioners often did not know how to conduct validation tasks. However, during practical exercises, they discovered tools like Wrtn and Bing AI, which provide source citations for answers, and Bard's "Google it" function, which allowed them to Google the generated answers to find similar content on the web. Subsequently, practitioners showed initiative in independently verifying the accuracy and truthfulness of answers. This highlighted the scattered nature of tools available for verifying AI responses, making it challenging for practitioners to find and use them individually. Furthermore, considering the opinion that a single verification process cannot verify all content, it is evident that there is a necessity to integrate multiple verification methods rather than relying on a single method. Therefore, integrating various verification methods such as source citation and web searches into a cohesive design can effectively meet practitioners' needs and preferences.

**Design Goal 3: A system that summarizes AI-generated responses and verification results to aid decision-making.**

One of the methods mentioned by participants as helpful for decision-making is the task of comparing multiple AI responses to find common content. This task, which involves comparing and analyzing large volumes of text, is complex. Therefore, a feature that can extract and compare common content among responses from various AIs could be beneficial. Furthermore, because AI generates large amounts of text, it becomes challenging to easily comprehend verification results when multiple verification methods are used concurrently. If the results of verification tasks can be synthesized into a single view, this could alleviate the complexity and difficulties associated with decision-making. Even junior practitioners, who were confused about which response to select among several in the absence of clear decision criteria, could benefit from being able to review organized results based on verification status. Hence, the third design goal is to create a system that summarizes AI-generated responses and verification results to aid decision-making.

## 4. Trustworthy AI Model

The results of the preliminary study showed that distrust toward AI-generated outputs in the initial two stages—problem definition (analyze and define) and solution generation (discover and ideation)—acts as the primary barrier to UXer-AI collaboration. One of the reasons for this distrust is the inability to verify the exact sources and evidence of the AI-generated outputs. Since it is currently technically challenging to fully explain an AI's working process, adding a separate verification process is a more realistic approach (Wang & Moulden, 2021) . Therefore, this study aims to expand the process by adding verification steps, focusing on the methods mentioned by workshop participants as effective.

### 4.1 Process Refinement: Four-Stage Process and Three-Mode AI Model

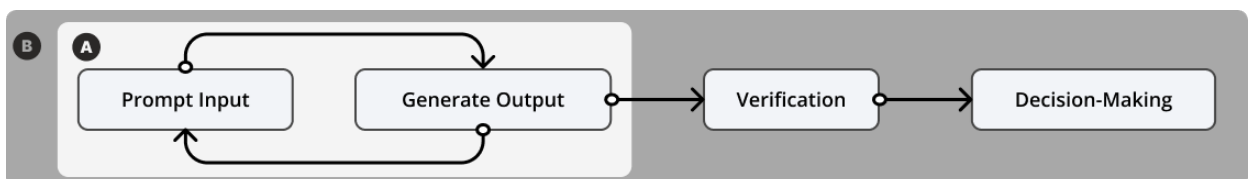

*Figure 1. A (2 stages). The initial UXer–AI collaboration process; B (4 stages). The newly proposed process.*

In the initial definition of the UXer–AI collaboration process, it was characterized by the repetition of two tasks: prompt input and output generation. Due to the lack of a separate verification process for responses, UXers could not trust the outputs generated by AI. There was also a risk of overlooking the necessity of verification and unquestioningly adopting AI-generated outputs. Therefore, this study expanded the existing UXer–AI collaboration process by adding response verification and decision-making stages (Figure 1), resulting in a process that iterates through four tasks. The flowchart depicting this process is shown in Figure 2 and comprises three modes: generation, verification, and decision-making.

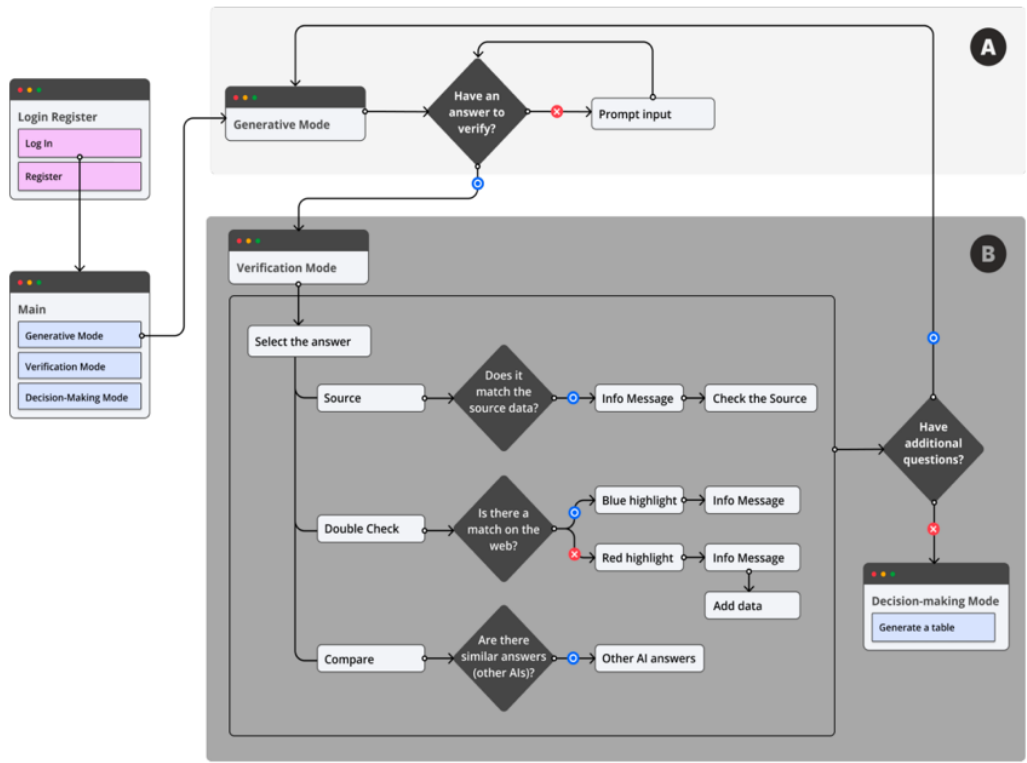

*Figure 2. Flow chart: The existing UXer–AI collaboration process merely repeats the cycle of UXers inputting prompts and AI generating responses (A). The new process allows for decision-making based on the results of verifying the generated responses through three different methods (B). Additionally, UXers with further questions can return to the initial stage, the generation mode.*

For convenience, the new AI prototype proposed in this paper will be referred to by the abbreviation TW-AI, which stands for "a new trustworthy AI model." The main screen of the TW-AI model, created according to the flow chart, features six key elements (Figure 3). First, on the left sidebar (❶), users can select one of the three modes: generation, verification, or decision-making. The default mode is generation, and users can switch modes according to their workflow using a button (❺). Consequently, the currently selected mode is displayed in the center of the screen (❸). At the top of the screen, there is a button (❷) users can use to access help and guidance for the TW-AI. This button opens a window providing detailed help for each mode. In the central part of the screen (❹), users can view quantitative service metrics necessary for UX research tasks, as well as a library of bookmarked conversations and prompts. At the bottom of the screen, there is an input field (❻) where users can enter prompts.

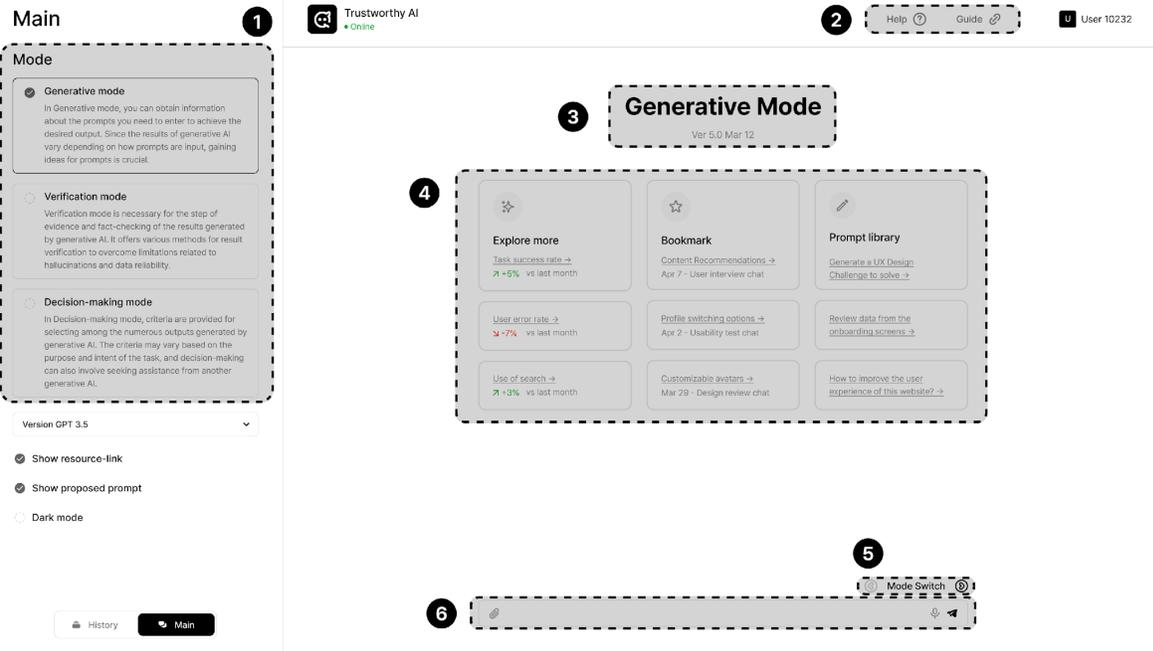

*Figure 3. Trustworthy AI (TW-AI) model: Full screen (with three mode menus on the left)*

The operation of the three modes is as follows. First, in the generation mode, prompts are entered and responses are generated, similar to existing generative AI systems. Next, in the verification mode, users select the response they wish to verify and perform three verification tasks on it (Figure 4). The details of these three verification tasks will be discussed in the following section.

Finally, in the decision-making mode, users can view the responses organized based on their reliability, which is determined by the results of the verification tasks (Figure 5). The criteria for assessing reliability are based on the three functions performed in verification mode: source, double-check, and compare, with each weighted accordingly. Responses that meet all three verification criteria are ranked highest while those partially verified are prioritized based on the extent of their verification. In UX design practice, making decisions based on research data is crucial. Therefore, providing UXers with summarized and verified responses can aid in making persuasive and evidence-based decisions.

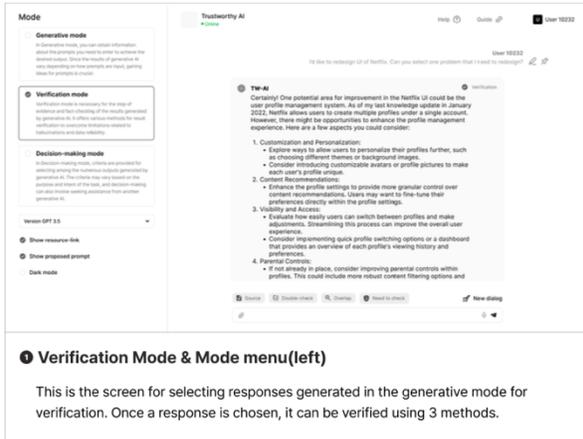
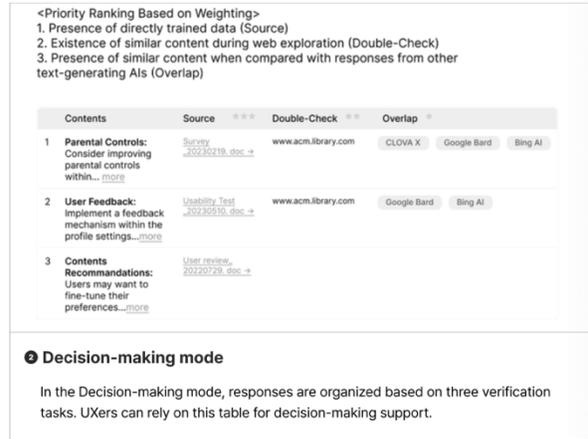

*Figure 4. 1) Verification mode; 2) Decision-making mode: Organizing answers in order of reliability based on verification work.*

## 4.2 Three Verification Functions

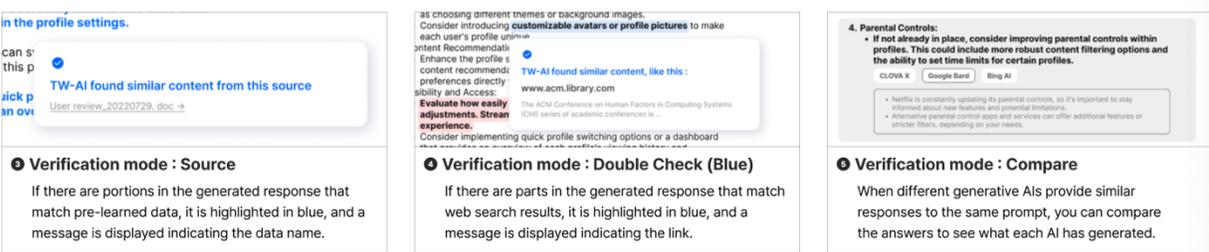

*Figure 5. 3) Guidance message for answers verified by source; 4) Guidance message for answers verified by double check; 5) Various AI responses viewable through the compare feature.*

### 4.2.1 Source: A Method to Verify If There Is a Match with the Data Directly Trained by the User

The source feature is a functionality that partially applies retrieval-augmented generation (RAG), allowing users to compare data they have directly trained into AI with AI-generated responses to identify matching content. Since each company has different target users and services, it is often necessary to use internal company data when engaging in UX design in practice. These data can include the results of internal usability tests or user research conducted by a company. Therefore, if an AI's responses can be compared with these data, the utilization of AI in practical UX design work would increase significantly. Workshop participants expressed the need for evidence to support AI responses, noting the importance of sources when writing reports in practice. They positively evaluated the functions of tools like Wrtn and Bing AI, which provide sources from the internet, and suggested that providing sources based on internal company data would have an even greater effect on improving trust.

### 4.2.2 Double Check: Verifying Consistency Between AI Responses and Web Content

Google Bard provides a Google it feature that shows whether AI-generated responses match content available on the web. Workshop participants positively evaluated this feature, noting its utility in obtaining evidence and discerning whether responses are hallucinations, thereby aiding in decision-making. Therefore, the second function of the verification mode, double check, adopted a similar approach to Google it. It compares AI responses with content on the web to identify similarities. If similar content is found on the web, this content is highlighted in blue, and double check provides relevant links. If no similar content is found, double check highlights this content in red and recommends further searches. If there are insufficient data to determine its truthfulness or if verification is unnecessary, no highlights are displayed.

### 4.2.3 Compare: A Method of Comparing Answers from Various AI Tools (Assistance in Decision-Making)

Text-generation AIs produce multiple responses to a single prompt. For instance, when a UXer input a prompt to identify issues in Netflix's UI, an AI generated various types of issues, requiring the UXer to identify and decide on the most critical ones. Therefore, without clear criteria or evidence for responses, decision-making was challenging for UXers. During the workshop, participants voiced such difficulties. Some participants addressed these challenges by comparing responses generated by different text-generation AIs after entering identical prompts. One participant highlighted that using both GPT and Bard emphasized the same issue, making it easier to define problems and make decisions (P18). Currently, accessing multiple AIs individually to input prompts and check responses is cumbersome and repetitive. However, the compare feature offers the advantage of directly comparing responses from multiple AIs with a single prompt input. Therefore, utilizing this feature would facilitate smoother decision-making processes for UXers.

## 5. EVALUATION
In this chapter, we validate the effectiveness of the TW-AI model implemented to address obstacles in UXer-AI collaboration. Existing studies have presented various scales for assessing AI trustworthiness (Hoffman et al., 2023; Israelsen & Ahmed, 2019; Ribeiro et al., 2016). However, trust is a multidisciplinary and complex concept, making it challenging to measure, and there is still no formally recognized trust model in use. The purpose of this study is to enhance the practical application of AI, which is why we applied the enterprise AI trustworthiness measurement model proposed by Wang and Moulden (2021). This model was designed to assess the experience of using AI in enterprise environments, making it suitable for assessing the extent to which UXers can trust and use AI in practical settings. In addition, meta-analysis has shown that scores for each measurement item predict user satisfaction with an AI's functionality.

The standard evaluation criteria of the AI trust measurement model (Wang & Moulden, 2021) include factors such as improving operational efficiency, understanding the timing and methods of use, controlling use, and security-related issues. These criteria are rated on a scale featuring three levels: good rating, okay rating, and needs improvement. Considering the research objectives, additional evaluation items from the previous study (Wang & Moulden, 2021) were added to the standard criteria to assess trust and confidence. In addition, the TW-AI model was created under the assumption that AI systems operate within an enterprise context. This assumption led to the exclusion of items related to the protection of customer data from the evaluation criteria. Subjective items regarding opinions and quantitative scoring were included to understand overall user satisfaction. The final set of five criteria is summarized in Figure 6. We used this AI trust assessment model to measure and compare the trustworthiness of both existing generative AI systems and the prototype AI, thereby gaining insight into overall user satisfaction.

| AI Trust Score Items | [The AI feature] will help me do my job more efciently and efectively. | |
|---|---|---|
| | I understand how and when to use [the AI feature]. | |
| | I have control using [the AI feature]. | |
| | I am confident in the results made by [the AI feature]. | |
| | I trust the results made by [the AI feature]. | |
| | Overall, how satisfed are you with using [the AI feature]? | |

- Good rating
- Okay rating
- Need improvement

*Figure 6 AI Trust Score Items : Reconstructed from Wang & Moulden (2021)*

## 5.1 Participants

Recruitment advertisements were posted in online communities frequented by UXers, resulting in the recruitment of 20 participants who are currently engaged in UX design or have relevant experience. Specifically, they were engaged in roles such as UX research, service planning, UX/UI design and product design. These participants were selected to undertake tasks related to "problem identification and definition" in the UX design process using both existing generative AI and a newly created generative AI designed by the researchers. Individuals without knowledge of UX design were excluded from participation.

## 5.2 Study Procedure

The study was structured into three stages: preliminary interviews, task performance using the existing generative AI and the prototype (TW-AI model), and in-depth interviews (Table 5).

*Table 5. Study Procedure*

| Participants | 20 UX design professionals | | | | | |
|---|---|---|---|---|---|---|
| Steps | Recruitment and preliminary interview | Existing generative AI | | Prototype (TW-AI model) | | In-depth Interview |
| | | Task execution | Trust evaluation | Task execution | Trust evaluation | |
| Contents | Investigation of generative AI usage experience | 1. Prompt input and output generation 2. Verification 3. Decision-making | Enterprise AI trust measurement model | 1. Prompt input and output generation 2. Verification 3. Decision-making | Enterprise AI trust measurement model | Evaluation of existing AI, trust level, prototype, and questions related to future changes |

In the first stage of task performance, participants used the existing generative AI tool to identify UX problems in Netflix. This specific focus on Netflix was chosen to reduce participant choice burden and standardize input prompts for the second task using the TW-AI model. Detailed task instructions are outlined in Table 6.

*Table 6. Task Instructions*

| Generative AI Tool | ChatGPT, Google Bard, wrtn, Clova X, Bing AI |
|---|---|
| Task Instructions | Using at least two of the available tools, complete the following tasks. |
| Task 1 (Research) | Input the two provided prompts into each selected AI and identify a problem in the Netflix UI that requires redesign. |
| Provided Prompts | - I'd like to redesign the UI of Netflix. Can you select one problem that I need to redesign? And translate the answer to Korean.<br>- Tell me about the most critical problem of Netflix's UI. And translate the answer to Korean. |
| Task 2 (Verification) | Find evidence to support the generated responses (e.g., web searches, service verification, and follow-up questions to the AI). |
| Task 3 (Decision-Making) | Based on Task 1 and 2, select one problem to focus on. |

During task execution, participants were guided to choose at least two of the following generative AI tools: ChatGPT, Google Bard, Wrtn, Naver Clova X, and Bing AI. This instruction was provided to facilitate the smooth completion of the second and third tasks, which involved verifying responses and making decisions. Task execution began with a participant inputting two provided prompts into the generative AI to generate responses. Then, they proceeded to verify the basis and accuracy of the generated responses freely by either questioning the AI again, conducting web searches, or directly accessing Netflix without any restrictions on methods.

After completing the validation process, they proceeded to the task of defining one problem that can serve as the redesign objective. The decision criteria for this task could be based on the validation results from the second task or on whether multiple AI models commonly mentioned the problem.
Following the completion of all tasks, a trustworthiness evaluation was conducted based on the AI trust scorecard. Participants provided ratings for five criteria and verbally articulated their overall satisfaction with the experience. They evaluated their usage experience by focusing on concerns in practical AI application and explained reasons for their scoring decisions.

In the second task-observation experiment, participants used the TW-AI model to perform the same three tasks as in the previous phase and evaluated the reliability scores. After completing all tasks, in-depth interviews were conducted to explore the UX of the TW-AI model, comparing its effectiveness with existing generative AI tools and identifying areas for improvement.

## 5.3 Data Analysis

The AI trust scores were calculated by assigning 1 point for a good rating, 0 points for an okay rating, and -1 points for a rating that needed improvement for each item. Although the number of study participants was not sufficient to derive statistical significance, the analysis was conducted in this way because it is meaningful to compare the trust in and satisfaction with the existing AI tools versus the TW-AI model based on the results of the in-depth interviews. The calculated average scores were plotted based on the five evaluation items shown in Figure 7. The overall average score for the existing AI tools was -0.1 while the overall average score for the TW-AI model was 3.65, indicating improvements across all items. The detailed changes in the scores for each trust evaluation item are explained in the next section.

## 6. RESULTS
### 6.1 Comparison of Trust Score Measurements

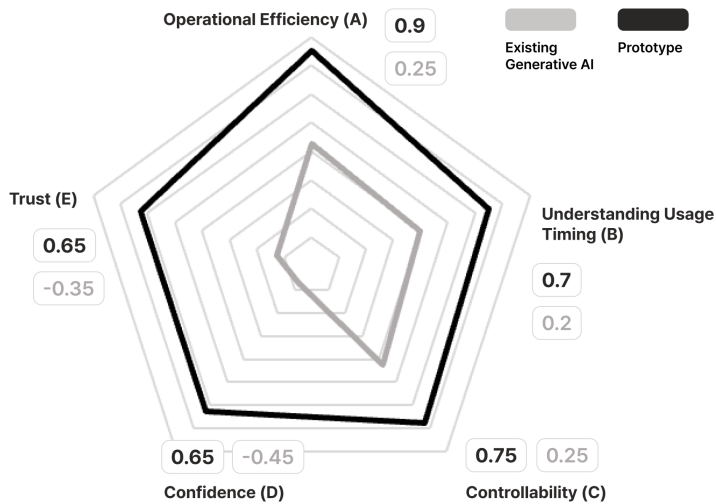

*Figure 7 AI trust scores*

### 6.1.1 Changes in Scores for the Confidence (D) and Trust (E) Items

Looking at the individual scores of the trust scorecard for the existing AI tools and the TW-AI model, the items confidence (D) and trust (E) showed the most significant improvements. In this context, confidence and trust refer to the degree of certainty and the reliability of the results generated by the AI, while overall trust, as indicated by the sum of the five detailed items, represents the trustworthiness of the AI tool itself.

Participants explained that the main problem with the existing AI tools is their inability to provide justifications, which made it difficult to trust, and they argued that the TW-AI model addressed this problem, thereby increasing their trust (P4, 6, 15, 16, 20). P7 mentioned that her confidence increased because the TW-AI performed verification tasks that she had previously done manually, resulting in final judgments that closely matched her own.

One participant stated that although the TW-AI model seemed more reliable, they still could not fully trust it. They elaborated, "Since I haven't directly experienced the results, I wonder if it's really correct." They added that repeated use of the TW-AI model may improve their trust over time (P18).

### 6.1.2 Changes in Scores for Operational Efficiency (A)

The overall increase in trust scores had a positive impact on the scores for the first item, which measures operational efficiency. Participants noted that higher levels of trust led to faster decision-making (P9) and allowed for multiple reviews, which saved significant time and therefore encouraged high efficiency scores (P12). In addition, participants (P1, 13) mentioned that the ability to trust information meant that it could be used more effectively, contributing to efficient task performance. P3 added that they could not use the existing AI tools in their work due to security concerns; however, the capability to customize the AI for each company, as proposed for the TW-AI model, increases its practical applicability in real-world scenarios.

### 6.1.3 Changes in Scores for Understanding Usage Timing (B) and Controllability (C)

Participants had different opinions on items B and C, which relate to understanding the timing and controllability of AI usage, respectively. First, participants who gave higher scores to the TW-AI model for understanding the timing of use mentioned that the existing AI tools did not clearly indicate when they should be used within the UX design process, whereas the TW-AI model allowed them to clearly recognize the timing (P8, P19). Specifically, P6 and P19 noted that the TW-AI model's three modes made it easier to understand the process. Regarding the controllability score for the TW-AI model, a participant who offered a high rating mentioned that the addition of a validation feature, which was previously lacking, made the control mechanism more convenient than before (P2). However, some participants rated the TW-AI model lower than the existing AI tools. Reasons for rating the understanding of timing lower included doubts about how to use the AI practically unless it has learned extensively about the service (P9) and concerns about overreliance on AI versus using it appropriately (P14). P4 rated the controllability of the TW-AI model lower, expressing that it felt more like following provided functionalities rather than directly controlling them, given that the TW-AI model is a prototype.

### 6.2 Analysis of In-Depth Interview Results
### 6.2.1 Causes of Trust Decline

The primary reason for the decline in trust was the lack of sources and evidence provided by traditional generative AI systems (P1, 10, 13, 15). P1 cited mistrust due to AI's inability to provide accurate information or base responses on user data. Hallucination was also identified as a cause of trust decline. Many participants refrained from using AI more frequently due to doubts about the answers themselves (P4, 5, 10, 16) and mistrust in AI tools (P1, 2). Additionally, exposure to news about AI-related hallucinations and outdated data updates led some participants to have lower baseline trust in AI (P12, 16), making it unusable in their eyes.

Second, the lack of a verification stage also decreased trust in AI. When using existing generative AI, a two-step process of user input followed by AI generation is repeated. Consequently, participants mentioned that they often overlooked verification tasks, stating, "I did not think to verify or find evidence while performing the task with AI responses" (P8, 9, 14). Particularly, junior practitioners with relatively less experience often failed to recognize the necessity of verification. Regarding this, P4 argued that because some people might skip verification, there should be a mandatory requirement for verification when collaborating with AI in practical work settings.

### 6.2.2 Factors Contributing to Improved Trust

First, the introduction of three verification features (source, double check, and compare) by the TW-AI model significantly enhanced trust in the generated data. Participants particularly valued the source feature, noting that generating and verifying responses based on internal company data results in much more objective and trustworthy data compared to human-generated data (P1, 10, 13, 19, 20). Moreover, the presence of multiple verification functions, rather than just one, further increased confidence in the final responses (P2, 5, 12). P5 explained, "Even if the double check result isn't available, I can still review it through the compare function, which increases my trust."

Second, the process of the TW-AI model, which includes verification and decision-making stages, improved trust in the AI tool. P1 pointed out that the existing AI tools generate answers without offering verification functions, making them seem indifferent to the accuracy of the information. She remarked, "The existing AI tools provide answers but don't offer verification functions, so it feels like they don't care whether the information is true or false. There is no conscious effort to ensure the user receives correct information." In contrast, she felt that the TW-AI model seemed to acknowledge the possibility of errors and made an effort to provide supporting evidence, thus presenting more responsible results. Similarly, P5 expressed that the provision of verification functions by the TW-AI model itself indicated a consideration for data reliability, which increased their trust in the tool.

### 6.2.3 Effectiveness of the Prototype

Beyond the improvement in trust, the most frequently mentioned benefit of the TW-AI model was time reduction (P1, 6, 8, 9, 10, 11, 12, 13, 14, 15, 18, 19). P8 described it as feeling like an "all-in-one" solution. P4, P17, and P18 highlighted that the research phase, which typically consumes the most time in the UX design process, was significantly shortened by the TW-AI model. Participants also anticipated that the TW-AI model could further enhance work efficiency by summarizing the service-related quantitative metrics or past research data (P7, 10, 19, 20).

Second, many participants found that the TW-AI model's decision-making mode aided in communication by providing evidence that can be presented to persuade others (P3, 4, 5, 7, 11, 18). P5 and P7 emphasized the increase in communication efficiency, noting that if the AI usage history could be shared with other employees, this would eliminate the need for separate documentation, allowing all communication to be conducted through the AI. Some participants also suggested that adding an image-generation feature could further enhance communication effectiveness (P5, 19, 20).

### 7. Discussion and Conclusion

This study highlighted that a major obstacle in UXer-AI collaboration is the issue of decreased trust, and it proposed an expanded UXer-AI collaboration process and an AI prototype model to address this challenge. The evaluation of the new AI prototype model revealed that AI tools supporting verification tasks and decision-making contribute to improved trust, offering insights into the design of AI tools specifically tailored for UX design tasks.

This study proposed a UXer-AI collaboration process and discussed effective strategies for collaborating with AI in practice. Initially, the UXer-AI collaboration process was defined through a literature review, followed by two participatory workshops with practitioners to identify obstacles in collaborating with AI. A strategy that includes verification and decision-making stages was proposed to mitigate these issues, thus providing insights into enhancing the essential component of trust in UXer-AI collaboration. The study offers the following implications.

### 7.1 Verification Functions for Enhancing AI Trust

UX design involves creating outputs based on research data, making trust a particularly critical factor in the success of UX design. In fact, the results of workshops with practitioners indicated that decreased trust is a major obstacle to UXer-AI collaboration, supporting prior research that identified trust as a key element in human-AI collaboration (Omrani et al., 2022; Ramchurn et al., 2021). In human-AI interactions in particular, trust can easily be lost as a result of initial experiences, making it crucial to establish cognitive trust early in the adoption of AI (Penny Collisson, 2025, as cited in Wang & Moulden, 2021). To address this, the study emphasized the need for verification tasks and proposed a UXer-AI collaboration process that includes fact-checking and decision-making, along with a prototype to implement this process. These solutions led to improved trust and efficiency by reducing the complexity and uncertainty central to trust (Muir, 1994). Among the three proposed verification functions, the source feature utilizing RAG technology was rated most highly, validating previous research that RAG can enhance AI accuracy and reliability and address hallucination issues (Gao et al., 2023).

### 7.2 The Role of UXers in Building Trustworthy UXer-AI Collaboration

The most important aspect of UXer-AI collaboration is for UXers to lead throughout the process. Despite AI providing verification functions and generating decision tables, humans remain responsible for conducting verification tasks and identifying the most critical issues from these tables. Research participants noted that while AI assists with verification and decision-making, human review and final decision-making are still necessary at every stage (P2, 4, 8, 12, 17). P12 expressed, "If trust in traditional AI is at 30%, then that for the TW-AI model is at 70%. For the remaining 30%, we still need to conduct research ourselves to verify." This underscores the crucial role of human review in credible UX research. Prior research also emphasized that reliable human-AI collaboration should leverage each party's strengths and address weaknesses (Ramchurn et al., 2021). By adopting this collaborative approach, potential risks, such as excessive dependence on AI, loss of human agency, and job displacement, can be mitigated (Buschek et al., 2020; Sison et al., 2023).

### 7.3 Expanding the Capabilities and Scopes of UXers

While most participants believed that the core role of designers would remain unchanged with AI collaboration, there was anticipation that the capabilities and scopes of UXers would expand (P1, 3, 5, 11). Participants mentioned that with AI collaboration, UXers would need to focus more on problem definition, direction setting, and decision-making rather than data collection and analysis (P1, 11, 14). Furthermore, the quality of AI responses can vary depending on the data an AI is trained on, highlighting the need to manage data separately (P3, 10, 14). Considering these opinions, the existing research suggesting that the potential for human designers to be replaced is low due to the existence of tasks AI cannot perform (P3, 10, 14) is likely applicable to the field of UX design as well.

### 7.4 Limitations and Future Directions

The rapid advancement of technologies, including generative AI, suggests that relying solely on current available technologies for UXer-AI collaboration strategies may present limitations. The prototype proposed in this study serves as an example of an AI tool for UXers that has the potential to evolve into various forms due to technological advancements and the additional needs of UXers. Furthermore, both the workshops and task observation experiments were conducted as one-time events. However, if UXer-AI collaboration were to occur repeatedly, issues distinct from those revealed by one-off workshops could arise. Therefore, future research could propose AI usage guidelines for UX design by conducting iterative collaborations with the same participants. Lastly, the prototype proposed in this study is applicable to the initial stages of UXer-AI collaboration processes, namely problem definition and solution generation. Validation methods tailored to image-generating AI or UI prototype-generating AI tools are necessary for the final stage, where the ultimate deliverables are produced. Future research is anticipated to expand and discuss validation methods when using image- and prototype-generating AI, thereby applying validation processes across the entire UX/UI design process.